\documentclass[12pt]{article}

\usepackage{geometry} 
\usepackage{graphicx} 

\usepackage{float} 
\usepackage{wrapfig} 
\usepackage{lipsum} 
\usepackage{exscale}
\usepackage{relsize}
\usepackage{indentfirst}
\usepackage{amssymb}
\usepackage{amsthm}
\usepackage{amssymb}
\usepackage{amsmath}
\usepackage{abstract}

\graphicspath{{./Pictures/}} 

\numberwithin{equation}{section}

\newtheorem{thm}{Theorem}[section]
\newtheorem{defn}[thm]{Definition}
\newtheorem{lem}[thm]{Lemma}
\newtheorem{coro}[thm]{Corollary}
\newtheorem{prop}[thm]{Proposition}
\newtheorem{remark}[thm]{Remark}
\newtheorem{exam}[thm]{Example}

\begin{document}
\title{Cyclic subspace codes via the sum of Sidon spaces\let\thefootnote\relax\footnotetext{
 E-Mail addresses: yunly@mails.ccnu.edu.cn (Y. Li), hwliu@mail.ccnu.edu.cn (H. Liu)}}
\author{Yun Li,~Hongwei Liu}
\date{\small School of Mathematics and Statistics, Central China Normal University, Wuhan, 430079, China \\}
\maketitle

\begin{abstract}
  Subspace codes, especially cyclic constant subspace codes, are of great use in random network coding.
  Subspace codes can be constructed by subspaces and subspace polynomials. In particular, many researchers are
  keen to find special subspaces and subspace polynomials to construct subspace codes with the size and
  the minimum distance as large as possible. In \cite{RRT}, Roth, Raviv and Tamo constructed several subspace codes
  using Sidon spaces, and it is proved that subspace codes constructed by Sidon spaces has the largest size and minimum distance.
  In \cite{NYW}, Niu, Yue and Wu extended some results of \cite{RRT} and obtained several new subspace codes.

  In this paper, we first provide a sufficient condition for the sum of Sidon spaces is again a Sidon space.
   Based on this result, we obtain new cyclic constant subspace codes through the sum of two and three Sidon spaces.
   Our results generalize the results in \cite{RRT} and \cite{NYW}.
\\


\end{abstract}

\section[Introduction]{Introduction}

Let $\mathbb{F}_q$ be a finite field with $q$ elements, and $\mathbb{F}_{q^n}$ be an
extension field of degree $n$ over $\mathbb{F}_q$. The field $\mathbb{F}_{q^n}$ can be viewed
as an $\mathbb{F}_q$-vector space with dimension $n$. Let $\mathcal{P}_q (n)$ denote the set
of all subspaces of $\mathbb{F}_{q^n}$, and let $\mathcal{G}_q(n,k)$ denote the set of
 subspaces of $\mathbb{F}_{q^n}$ with dimension $k$. A {\it subspace code} $\mathcal{C}$ is a nonempty subset of
$\mathcal{P}_q (n)$. The distance of $\mathcal{P}_q (n)$ is defined as\\
\centerline{$d(U,V)=\dim U+\dim V-2\dim (U \cap V)$,}
where $U, V \in \mathcal{P}_q (n)$. The {\it minimum distance of the subspace code} $\mathcal{C}$ is defined
as\\
\centerline{$d(\mathcal{C})=\textup{min} \{d(U,V)\mid U\neq V, U,V\in\mathcal{C}$\}.}
If for all subspaces contained in $\mathcal{C}$ has the same dimension, then
 $\mathcal{C}$ is called a {\it constant dimension subspace code}.
The subspace code $\mathcal{C}$ is called {\it cyclic} if for all nonzero $\alpha\in\mathbb{F}_{q^n}$
and $U\in\mathcal{C}$, we have $\alpha U\in \mathcal{C}$.  Let $V \in \mathcal{P}_q (n)$, the multiplicative group $\mathbb{F}_{q^n}^*$
acts transitively on the set $\textup{orb}(V)=\{\alpha V\mid \alpha\in\mathbb{F}_{q^n}^*\}$.
It is easy to verify that $\textup{orb}(V)$ is a cyclic constant dimension subspace code.
The size of $\mathcal{C}=\textup{orb}(V)$ is equals to $\frac{q^n-1}{q^t-1}$,
 where $t$ is the least positive integer such that $V$ is also an $\mathbb{F}_{q^t}$-vector space, and the minimum distance of $\mathcal{C}=\textup{orb}(V)$ is equal to
 $2k-2s$, where $s$ is an integer with $0\leq s\leq k$.
The best minimum distance of a code $\mathcal{C}$ whose size is greater than or equal to $\frac{q^n-1}{q-1}$ can attain is $2k-2$.
 Therefore, many researchers focus on constructing subspace codes with size $\frac{q^n-1}{q-1}$ and
 minimum distance $2k-2$. Subspace codes are constructed in this way mentioned below.

In \cite{ACLY}, Ahlswede et.al. introduced network coding.
 K\"oetter and Kschischang introduced subspace codes, and they constructed
constant dimension codes of projective space in \cite{KK}. Subspace codes are of great use in random network coding.
Ben-Sasson et.al. constructed
subspace codes with size $\frac{q^n-1}{q-1}$ and minimum distance $2k-2$ by the subspace
polynomial $T(x)=x^{q^k}+x^{q}+x$ in \cite{BEGR}. They combined codes constructed by Frobenius substitution to enlarge the size of subspace codes.
The size of the code combined by $r$ subspace codes is $r\frac{q^n-1}{q-1}$, and the minimum distance is still $2k-2$.
In \cite{OO}, Otal and \"Ozbudak also constructed subspace codes by combining several subspace codes
via $r$ different subspace polynomials $T(x)=x^{q^k}+\theta_i x^q+\gamma_i x$ with different coefficients.
Chen and Liu constructed subspace codes in \cite{CL} by subspace polynomials
$T(x)=x^{q^k}+a_l x^{q^l}+ a_0 x$ which satisfy the condition $\gcd(k,l)=1$. In this case,
sizes and minimum distances of subspace codes via subspace polynomials $T(x)=x^{q^k}+a_l x^{q^l}+ a_0 x$ are also $\frac{q^n-1}{q-1}$ and $2k-2$.

Recently, Roth, Raviv and Tamo \cite{RRT} provided new methods of constructing subspace codes using Sidon spaces.
It is proved that the cyclic constant subspace code $\textup{orb}(V)$ has size $\frac{q^n-1}{q-1}$ and minimum distance
$2k-2$ if and only if $V$ is a Sidon space. They also
constructed Sidon space $V=\{u+u^q\gamma\mid u\in\mathbb{F}_{q^k}$\}, where $\gamma$ is a root of an irreducible polynomial of degree $\frac{n}{k}>2$
over $\mathbb{F}_{q^k}$.
Niu, Yue and Wu constructed new Sidon spaces in \cite{NYW}. They constructed Sidon subspace
$V=\{a+u\gamma+u^q\gamma^2 \mid u\in\mathbb{F}_{q^k},a\in\mathbb{F}_q\}$, where $\gamma$ is a root of an irreducible polynomial of degree $\frac{n}{k}>4$ over $\mathbb{F}_{q^k}$, and
$V=\{a+a^q\gamma+u\gamma^2+u^q\gamma^3 \mid u\in\mathbb{F}_{q^l},a\in\mathbb{F}_{q^k}\}$, where
$\gamma$ is a root of an irreducible polynomial of degree $\frac{n}{k}>6$ over $\mathbb{F}_{q^{kl}}$ and $\gcd(k,l)=1$.
Then new subspace codes are constructed through these Sidon spaces.

In this paper,  our goal is to construct cyclic constant subspace codes through Sidon spaces that are sum of several Sidon spaces.
This paper is organized as follows.
In Section 2,
we give some notations and review some basic results that will be used in subsequent sections.
In Section 3,
we give a sufficient condition for the sum of some Sidon spaces to be a Sidon space.
In Section 4,
we construct subspace codes through the sum of two and three Sidon spaces satisfy the sufficient condition given in Section 3.
We also construct subspace codes by combining distinct subspace codes.
Section 5 concludes our results.

\section{Preliminaries}
Recall that $\mathbb{F}_{q^n}$ is an extension field over $\mathbb{F}_q$, and it also can be viewed as an
$\mathbb{F}_q$-vector space with dimension $n$. Let $\mathcal{G}_q(n,k)$ denote the set of subspaces of $\mathbb{F}_{q^n}$
with dimension $k$.
Let $U,V$ be subspaces of $\mathbb{F}_{q^n},$ we denote $U+V$ to be the sum of $U$ and $V$ as subspaces, and denote $U\oplus V$
to be the direct sum of $U$ and $V$ as subspaces.
Let $UV$ be the $\mathbb{F}_q$-linear span of the set of products $uv$, where $u\in U, v\in V$.
And recall that \textup{orb}$(V)=\{\alpha V \mid \alpha\in\mathbb{F}_{q^n}^*\}$ is a cyclic constant subspace code.

\begin{defn}\label{def2.1}
  \cite{BSZ} A subspace $V\in\mathcal{G}_q(n,k)$ is called a Sidon space if for any nonzero elements $a,b,c,d \in V$,
  if $ab=cd$, then $\{a \mathbb{F}_q, b \mathbb{F}_q\}=\{c \mathbb{F}_q, d \mathbb{F}_q\}$.

\end{defn}


\begin{prop}\label{prop2.3}
  \cite{RRT}
   For a subspace $V \in  \mathcal{G}_q(n,k)$, the code $orb(V)$ is of size $\frac{q^n-1}{q-1}$
  and minimum distance $2k-2$ if and only if $V$ is a Sidon space.
\end{prop}

\begin{prop}\label{prop2.4}
  \cite{RRT}
  The following two conditions are equivalent for any distinct subspaces $U$ and $V$ in $\mathcal{G}_q(n,k)$.\\
  (1) $\dim(U\cap \alpha V)\leq 1,$ for any $\alpha\in \mathbb{F}_{q^n}^*.$\\
  (2) For any nonzero $a,c\in U $ and nonzero $b,d\in V$, the equality $ab=cd$ implies that $\{a \mathbb{F}_q\}=\{c \mathbb{F}_q\}$
   and $\{b \mathbb{F}_q\}=\{d \mathbb{F}_q\}$.
\end{prop}

The above two propositions are very useful in this paper. Proposition 2.3 implies  that we can construct
subspace codes with larger size and larger minimum distance by constructing Sidon spaces.
Proposition \ref{prop2.4} will be frequently used in Section 3 and 4.

Let $W_{q-1}=\{x^{q-1}\mid x\in \mathbb{F}_{q^k}\}$ and $\overline{W}_{q-1}=\mathbb{F}_{q^k}\backslash{W}_{q-1}$.
We have the following proposition.
\begin{prop}\label{prop2.5}
  \cite{RRT}
  For a prime power $q\geq3$ and a positive integer $k$, let $n=2k$, let $\gamma\in\mathbb{F}_{q^n}^*$
  be a root of an irreducible polynomial $x^2+bx+c$ over $\mathbb{F}_{q^k}$ with $c\in\overline{W}_{q-1}$,
 then $\mathcal{U}=\{u+u^q \gamma \mid u\in \mathbb{F}_{q^k}\}$ is a Sidon space.

\end{prop}

\section{The sum of Sidon spaces}

In this section, we give a sufficient condition such that the sum of several Sidon spaces is again a Sidon space.

\begin{lem}\label{lem3.1}
  Let $U,V\leq\mathbb{F}_{q^n}$ be two Sidon spaces. If $U$ and $V$ satisfy the following conditions, \\
  (1) $(U+V)^2=U^2\oplus UV\oplus V^2$,\\
  (2) $\dim(U\cap \alpha V)\leq 1$ for any $\alpha\in \mathbb{F}_{q^n}^*,$\\
  then $U+V$ is also a Sidon space.
  \begin{proof}
    For any nonzero elements $\alpha =a+b, \beta=c+d,\alpha' =a'+b', \beta'=c'+d'$ in $U+V$, where $ a,a',c,c'\in U $, and $ b,b',d,d'\in V$.
    If $\alpha\beta=\alpha'\beta'$, then we have\\
    \centerline{$ac+ad+bc+bd=a'c'+a'd'+b'c'+b'd'.$}
  By (1), this equality implies that
  \begin{equation}\label{gs3.1}
    \left\{
      \begin{aligned}
      ac&=a'c',\\
      bd&=b'd',\\
      ad+bc&=a'd'+b'c'.
    \end{aligned}
    \right.
    \end{equation}

    In the following, we prove the lemma in two cases.

    Case $1: abcd\neq 0$. Since $U$ is a Sidon space by the first equality in Equation (\ref{gs3.1}), we have
$\{a\mathbb{F}_q,c\mathbb{F}_q\}=\{a'\mathbb{F}_q,c'\mathbb{F}_q\}$. This implies that
$\frac{a}{a'}=\frac{c'}{c}\in\mathbb{F}_q^*$ or $\frac{a}{c'}=\frac{a'}{c}\in\mathbb{F}_q^*.$
Suppose
$\frac{a}{a'}=\frac{c'}{c}=\lambda\in\mathbb{F}_q^*, \frac{b}{b'}=\frac{d'}{d}=\delta\in\mathbb{F}_{q^n}^*$.

 (\romannumeral1) If $\lambda=\delta,$ then we have $\frac{\alpha}{\alpha'}=\frac{\beta'}{\beta}=\lambda\in\mathbb{F}_q^*.$

 (\romannumeral2) If $\lambda\neq\delta,$ then the third equality in Equation (\ref{gs3.1}) can be changed to \\
\centerline{$\lambda a'd+\delta b'c=\delta a'd+\lambda b'c,$}
which means \\
\centerline{$a'd=b'c$.}
By (2) and Proposition \ref{prop2.4}, we have $\frac{a}{c'}=\frac{a'}{c}=\frac{b'}{d}=\frac{b}{d'}=\tau\in\mathbb{F}_q^*.$
Therefore, $\frac{\alpha'}{\beta}=\frac{\alpha}{\beta'}=\tau\in\mathbb{F}_q^*.$

 It can be proved similarly when $\frac{a}{c'}=\frac{a'}{c}\in\mathbb{F}_q^*$.




 Case $2: abcd= 0$.

 (\romannumeral1) If only one member of $\{a,b,c,d\}$ is zero.
 Without loss of generality, we can assume that $a=0,$ then we have $a'=0$ or $c'=0$.

 If $a'=0$, it is easy to obtain that
 $\frac{b}{b'}=\frac{c'}{c}=\frac{d'}{d}\in\mathbb{F}_{q}^*.$ Thus, we have $\frac{\alpha}{\alpha'}=\frac{\beta'}{\beta}\in\mathbb{F}_{q}^*.$

 If $c'=0$, it is easy to obtain that $\frac{b}{d'}=\frac{a'}{c}=\frac{b'}{d}\in\mathbb{F}_{q}^*.$ Thus, we have
 $\frac{\alpha'}{\beta}=\frac{\alpha}{\beta'}\in\mathbb{F}_{q}^*.$

(\romannumeral2) If exactly two members of $\{a,b,c,d\}$ are equal to zeroes.
By assumptions of $\alpha$ and $\beta$, we have $a=c=0$, $a=d=0$, $b=c=0$ or $b=d=0$.

If $a=c=0,$ it is easy to verify that $a'=c'=0$. Then we have $\frac{\alpha}{\alpha'}=\frac{b}{b'}\in\mathbb{F}_q^*$
or $\frac{\alpha}{\beta'}=\frac{b}{d'}\in\mathbb{F}_q^*.$

If $a=d=0$, it is easy to verify that $a'=d'=0$ or $c'=b'=0.$ If $a'=d'=0$, then we have
$\frac{\alpha}{\alpha'}=\frac{b}{b'}\in\mathbb{F}_q^*$.
If $c'=b'=0$, we have $\frac{\alpha}{\beta'}=\frac{b}{d'}\in\mathbb{F}_q^*$.

Other two cases are similar.
  \end{proof}
\end{lem}

In the following theorem, we discuss the property of the sum of $m$ Sidon spaces of $\mathbb{F}_{q^n}$.

\begin{thm}\label{thm3.2}
  Let $V_i\leq\mathbb{F}_{q^n}$ be Sidon spaces, where $i=1,\cdots,m$ and $m$ is a positive integer with $m\geq 2$. If $V_1,\cdots,V_m$ satisfy the following conditions, \\
  (1) $\left(\displaystyle\sum_{i=1}^{m} V_i\right)^2=\displaystyle\bigoplus_{1\leq i\leq j\leq m}^{} V_i V_j$,\\
  (2) $\dim(V_i\cap \alpha V_j)\leq 1$, for any $1\leq i\neq j\leq m$ and $\alpha\in \mathbb{F}_{q^n}^*,$\\
  then $\displaystyle\sum_{i=1}^{m} V_i$ is also a Sidon space.

  \begin{proof}

    We prove it by induction. If $m=2,$ then $V_1+V_2$ is a Sidon space by Lemma \ref{lem3.1}.
    Now we assume that the conclusion is true when $m=k,$ where $k\geq 2$ is a positive integer.
    It is enough to prove that it is true when $m=k+1$. Since\\
    \centerline{$\displaystyle\sum_{i=1}^{k+1} V_i=\displaystyle\sum_{i=1}^{k} V_i +V_{k+1},$}
    and we know that $\displaystyle\sum_{i=1}^{k} V_i$ is a Sidon space through our hypothesis,
    then we have \\
      \begin{align*}
        \left(\displaystyle\sum_{i=1}^{k+1} V_i\right)^2&=\left(\displaystyle\sum_{i=1}^{k} V_i+V_{k+1}\right)^2
        =\displaystyle\bigoplus_{1\leq i\leq j\leq k+1}^{} V_i V_j\\
        &=\left(\displaystyle\bigoplus_{1\leq i\leq j\leq k}^{} V_i V_j\right)\oplus \displaystyle\sum_{i=1}^{k} V_i V_{k+1}\oplus V_{k+1}^2\\
        &=\left(\displaystyle\sum_{i=1}^{k} V_i\right)^2\oplus \left(\displaystyle\sum_{i=1}^{k} V_i\right) V_{k+1}\oplus V_{k+1}^2.
    \end{align*}

    For any nonzero elements $a=\displaystyle\sum_{i=1}^{k} a_i, a'=\displaystyle\sum_{i=1}^{k} a_i'\in \displaystyle\sum_{i=1}^{k} V_i,$ where $a_i, a_i'\in V_i,$ and nonzero elements $b, b'\in V_{k+1}$.
  If $ab=a'b'$, we have $a_ib=a_i'b'$. This implies $\frac{b'}{b}=\frac{a_i}{a_i'}\in\mathbb{F}_q^*$ or $a_i=a_i'=0,i=1,\cdots,k.$ Therefore, $\frac{b'}{b}=\frac{a}{a'}\in\mathbb{F}_{q}^*$.
  From the Proposition \ref{prop2.4}, we know that $\dim\left(\displaystyle\sum_{i=1}^{k} V_i, \alpha V_{k+1} \right)\leq 1,$
  for any $\alpha\in\mathbb{F}_{q^n}^*.$ By Lemma \ref{lem3.1}, we draw the conclusion that $\displaystyle\sum_{i=1}^{k+1} V_i$
  is a Sidon space.
  \end{proof}
\end{thm}

\section{Construction of cyclic constant subspace codes}
In \cite{RRT}, it is proved that the cyclic constant subspace code $\textup{orb}(V)$ has size $\frac{q^n-1}{q-1}$ and minimum distance
$2k-2$ if and only if $V$ is a Sidon space. Now we construct Sidon spaces
which are the sum of some Sidon spaces.

\subsection{Constructing Sidon spaces through two Sidon spaces}

\begin{lem}\label{lem4.1}

  For two positive integers $n,k$ and $k\mid n.$ Let $\gamma\in\mathbb{F}_{q^n}^*$ be a root of an irreducible polynomial of degree $\frac{n}{k}> 2$
  over $\mathbb{F}_{q^{k}}$. Let \\
  \centerline{$U=\{\delta u+\tau u^q\gamma \mid u\in \mathbb{F}_{q^k}\}$,}
  where $\delta,\tau \in\mathbb{F}_{q^k}^*$. Then $U$
  is a Sidon space with dimension $k$.

\begin{proof}
  For any nonzero elements $\alpha=\delta u+\tau u^q\gamma,\beta=\delta v+\tau v^q\gamma,\alpha'=\delta u'+\tau u'^q\gamma,\beta'=\delta v'+\tau v'^q\gamma\in U$,
  we have\\
  \centerline{$\alpha\beta=(\delta u+\tau u^q\gamma)(\delta v+\tau v^q\gamma)=\delta^2uv+\delta\tau( uv^q+ vu^q)\gamma+\tau^2 u^qv^q\gamma^2.$}
  Since $\frac{n}{k}>2$, we know that $\{1,\gamma,\gamma^2\}$ are linearly independent over $\mathbb{F}_{q^k}$. Let\\
  \centerline{$f(x)=\delta^2uv+\delta\tau(uv^q+ vu^q)x+\tau^2 u^qv^qx^2\in\mathbb{F}_{q^k}[x]$}
   and \\
  \centerline{$f'(x)=\delta^2u'v'+\delta\tau(u'v'^q+v'u'^q)x+\tau^2 u'^qv'^qx^2\in\mathbb{F}_{q^k}[x].$}
  If $\alpha\beta=\alpha'\beta'$, then $f(x)=f'(x).$ Therefore, they have the same roots.
  We suppose the root $\frac{-\delta u}{\tau u^q}=\frac{-\delta u'}{\tau u'^q},$ then $(\frac{u}{u'})^{q-1}=1,$ thus $\frac{u}{u'}\in\mathbb{F}_{q}^*.$
  This means that $\frac{\alpha}{\alpha'}=\frac{\beta'}{\beta}\in\mathbb{F}_{q}^*.$ Other cases are similar.
\end{proof}

\end{lem}

  \begin{lem}\label{lem4.2}
    For four positive integers $n,k,t,s$ and $k\mid n,\gcd(t-s,k)=1.$
    Let $\gamma\in\mathbb{F}_{q^n}^*$ be a root of an irreducible polynomial of degree $\frac{n}{k}> 2$
    over $\mathbb{F}_{q^{k}}$. Then \\
    \centerline{$U=\{u^{q^t}+u^{q^s}\gamma \mid u\in \mathbb{F}_{q^k}\}$}
    is a Sidon space with dimension $k$.

    \begin{proof}
      For any nonzero elements \\
      \centerline{$\alpha=u^{q^t}+u^{q^s}\gamma,\beta=v^{q^t}+v^{q^s}\gamma,\alpha'=u'^{q^t}+u'^{q^s}\gamma,\beta'=v'^{q^t}+v'^{q^s}\gamma\in U$,}
      \centerline{$\alpha\beta=u^{q^t}v^{q^t}+(u^{q^t}v^{q^s}+u^{q^s}v^{q^t})\gamma+u^{q^s}v^{q^s}\gamma^2.$}
      Since $\frac{n}{k}>2$, we know that $\{1,\gamma,\gamma^2\}$ are linearly independent over $\mathbb{F}_{q^k}$. Let \\
      \centerline{$f(x)=u^{q^t}v^{q^t}+(u^{q^t}v^{q^s}+u^{q^s}v^{q^t})x+u^{q^s}v^{q^s}x^2\in\mathbb{F}_{q^k}[x]$}
       and \\
      \centerline{$f'(x)=u'^{q^t}v'^{q^t}+(u'^{q^t}v'^{q^s}+u'^{q^s}v'^{q^t})x+u'^{q^s}v'^{q^s}x^2\in\mathbb{F}_{q^k}[x].$}
      This means that if $\alpha\beta=\alpha'\beta'$, then $f(x)=f'(x),$ thus they have the same roots.
      We suppose the root $\frac{- u^{q^t}}{u^{q^s}}=\frac{- u'^{q^t}}{u'^{q^s}},$ then $(\frac{u}{u'})^{q^t-q^s}=1.$
      Thus, $(\frac{u}{u'})^{q^s(q^{t-s}-1)}=1$, since $\gcd(t-s,k)=1,$ we have $\frac{u}{u'}\in\mathbb{F}_{q}^*.$
      This means that $\frac{\alpha}{\alpha'}=\frac{\beta'}{\beta}\in\mathbb{F}_{q}^*.$ Other cases are similar.
    \end{proof}
  \end{lem}

  By taking $\delta=\tau=1$ in Lemma \ref{lem4.1}, or $t=0, s=1$ in Lemma \ref{lem4.2}, we get Theorem 16 in \cite{RRT}.

  \begin{coro}([14], Theorem 16)
    For two positive integers $n,k$ and $k\mid n.$ Let $\gamma\in\mathbb{F}_{q^n}^*$ be a root of an irreducible polynomial of degree $\frac{n}{k}> 2$
    over $\mathbb{F}_{q^{k}}$. Then\\
    \centerline{$U=\{u+u^q\gamma \mid u\in \mathbb{F}_{q^k}\}$}
   is a Sidon space with dimension $k$.
    \end{coro}

\begin{remark}
 In Lemma \ref{lem4.1},
 if we change $u$ and $u^q$ to $u^{q^t}$ and $u^{q^s}$, respectively, where $\gcd(t-s,k)=1$,
  then $U$ is also a Sidon space. Based on this result, all Sidon spaces constructed
  in this paper can be changed in this way to obtain some other Sidon spaces.

\end{remark}

It is easy to verify that if $U$ is a Sidon space of $\mathbb{F}_{q^n}$ then $\alpha U$ is still
a Sidon space for any nonzero element $\alpha\in\mathbb{F}_{q^n}$.
In the following theorem, we construct a Sidon space which is the sum of $\gamma U$
and $\mathbb{F}_q$, where $U$ is a Sidon space constructed in Lemma \ref{lem4.1} and
$\mathbb{F}_q$ is trivial Sidon space of $\mathbb{F}_{q^n}$.

\begin{thm}\label{thm4.5}
  For two positive integer $n,k$ and $k\mid n.$
  Let $\gamma\in\mathbb{F}_{q^n}^*$ be a root of an irreducible polynomial of degree $\frac{n}{k}> 4$
  over $\mathbb{F}_{q^{k}}$. Let \\
  \centerline{$\mathcal{U}=\{a+\delta_1u\gamma+\delta_2u^q\gamma^2\mid a\in \mathbb{F}_{q},u\in \mathbb{F}_{q^k}\}$,}
  where $\delta_i\in\mathbb{F}_{q^k}^*, i=1,2.$
  Then $\mathcal{U}$ is a Sidon space with dimension $k+1$.

  \begin{proof}
    We assume that $U=\{\delta_1 u\gamma+\delta_2 u^q\gamma^2\mid u\in\mathbb{F}_{q^k}\}.$ Then $U$ is a Sidon space.
 Firstly, we prove that these two Sidon spaces $\mathbb{F}_q,U$ satisfy (1) in Lemma \ref{lem3.1}.
 Let
 \begin{center}
  $\begin{aligned}
    \alpha_1&=a\in\mathbb{F}_q,\\
    \alpha_2&=\delta_1 u\gamma+\delta_2 u^q\gamma^2\in U,\\
    \alpha_3&=\displaystyle\sum_{i = 1}^{m}\lambda_{i}[\delta_1^2v_iw_i\gamma^2+\delta_1\delta_2(v_i^qw_i+v_iw_i^q)\gamma^3+\delta_2^2v_i^qw_i^q\gamma^4]\in U^2,
  \end{aligned}$
 \end{center}
where $a,\lambda_{i}\in\mathbb{F}_q, u,v_i,w_i\in \mathbb{F}_{q^k}.$

Since $\frac{n}{k}>4$, we know that $\{1,\gamma,\gamma^2,\gamma^3,\gamma^4\}$ are linearly independent over $\mathbb{F}_{q^k}$.
If $\alpha_1+\alpha_2+\alpha_3=0$, it is easy to know $\alpha_1=0.$ Then we have
 $\displaystyle\sum_{i = 1}^{m}\lambda_{i}\delta_2^2v_i^qw_i^q\gamma^4=0$ and $\displaystyle\sum_{i = 1}^{m}\lambda_{i}\delta_1\delta_2(v_i^qw_i+v_iw_i^q)\gamma^3=0$. Thus, $\displaystyle\sum_{i = 1}^{m}\lambda_{i}v_i^qw_i^q=0,$ this means
 $\displaystyle\sum_{i = 1}^{m}\lambda_{i}v_iw_i=0.$ Therefore, we have $\alpha_3=0$.
 Hence, $\alpha_1=\alpha_2=\alpha_3=0$. This implies
 that $U+U^2+\mathbb{F}_q=U\oplus U^2\oplus \mathbb{F}_q$.

Secondly, we prove that $U,\mathbb{F}_q$ satisfy (2) in Lemma \ref{lem3.1}.
For any nonzero elements $\delta_1 u\gamma+\delta_2 u^q\gamma^2, \delta_1 v\gamma+\delta_2 v^q\gamma^2\in U$ and $a,b\in \mathbb{F}_q^*.$
If\\
 \centerline{$a(\delta_1 u\gamma+\delta_2 u^q\gamma^2)=b(\delta_1 v\gamma+\delta_2 v^q\gamma^2)$,}
  then we have
$\delta_1 au=\delta_1 bv$, which means $au=bv.$ Hence, $\frac{u}{v}=\frac{b}{a}=\frac{\delta_1 u\gamma+\delta_2 u^q\gamma^2}{\delta_1 v\gamma+\delta_2 v^q\gamma^2}=\lambda\in\mathbb{F}_{q}^*.$

By Lemma \ref{lem3.1}, we conclude that $\mathcal{U}$ is a Sidon space.
  \end{proof}
\end{thm}

\begin{remark}([12], Theorem 3.1)
  By taking $\delta_1=\delta_2=1$, we get Theorem 3.1 in \cite{NYW}.
\end{remark}

Now we construct a Sidon space which is the sum of two nontrivial Sidon spaces.

\begin{thm}\label{thm4.7}
  For three positive integers $n,k,l$ and $kl\mid n,$ $\gcd(k,l)=1.$
  Let $\gamma\in\mathbb{F}_{q^n}^*$ be a root of an irreducible polynomial of degree $\frac{n}{kl}> 6$
  over $\mathbb{F}_{q^{kl}}$. Let \\
  \centerline{$\mathcal{U}=\{\delta_1 a+\tau_1 u\gamma+\delta_2 a^q\gamma^2+ \tau_2 u^q\gamma^3\mid a\in \mathbb{F}_{q^k},u\in \mathbb{F}_{q^l}\},$}
  where $\delta_i\in\mathbb{F}_{q^k}^*, \tau_i\in\mathbb{F}_{q^l}^*, i=1,2.$
  Then $\mathcal{U}$ is a Sidon space with dimension $k+l$.

  \begin{proof}
    We assume that $U=\{\delta_1 a+\delta_2 a^q\gamma^2\mid a\in \mathbb{F}_{q^k}\}, V=\{\tau_1 u\gamma+\tau_2 u^q\gamma^3\mid u\in \mathbb{F}_{q^l}\}$, then $U,V$ are all Sidon spaces.
    Firstly, we prove that $U,V$ satisfy (1) in Lemma \ref{lem3.1}.
    Let
   \begin{center}
    $\begin{aligned}
   \alpha_1&=\displaystyle\sum_{i = 1}^{m_1}\lambda_{1i}[\delta_1\tau_1 a_iu_i\gamma+(\delta_2\tau_1a_i^qu_i+\delta_1\tau_2a_i u_i^q)\gamma^3+\delta_2\tau_2a_i^qu_i^q\gamma^5]\in UV,\\
   \alpha_2&=\displaystyle\sum_{i = 1}^{m_2}\lambda_{2i}[\delta_1^2b_ic_i+\delta_1\delta_2(b_i^qc_i+b_ic_i^q)\gamma^2+\delta_2^2b_i^qc_i^q\gamma^4]\in U^2, \\
   \alpha_3&=\displaystyle\sum_{i = 1}^{m_3}\lambda_{3i}[\tau_1^2v_iw_i\gamma^2+\tau_1\tau_2(v_i^qw_i+v_iw_i^q)\gamma^4+\tau_2^2v_i^qw_i^q\gamma^6]\in V^2,
    \end{aligned}$
   \end{center}
   where $\lambda_{ij}\in\mathbb{F}_q, a_i,b_i,c_i\in \mathbb{F}_{q^k}, u_i,v_i,w_i\in\mathbb{F}_{q^l}.$

   Since $\frac{n}{kl}>6$, we know that $\{\gamma^i\mid 0\leq i\leq 6\}$ are linearly independent over $\mathbb{F}_{q^{kl}}$.
   If $\alpha_1+\alpha_2+\alpha_3=0$, it is easy to know $\alpha_1=0.$ Then we have
    $\displaystyle\sum_{i = 1}^{m_2}\lambda_{2i}\delta_1^2b_ic_i=0$, which leads to $\displaystyle\sum_{i = 1}^{m_2}\lambda_{2i}b_ic_i=0.$ Thus, $\displaystyle\sum_{i = 1}^{m_2}\lambda_{2i}^qb_i^qc_i^q=0,$ then
$\alpha_2=\displaystyle\sum_{i = 1}^{m_2}\lambda_{2i}\delta_1\delta_2(b_i^qc_i+b_ic_i^q)\gamma^2.$ Since
$\displaystyle\sum_{i = 1}^{m_3}\lambda_{3i}\tau_2^2v_i^qw_i^q\gamma^6=0,$ then $\displaystyle\sum_{i = 1}^{m_3}\lambda_{3i}v_i^qw_i^q=\displaystyle\sum_{i = 1}^{m_3}\lambda_{3i}^qv_i^qw_i^q=0,$ which
leads to $\displaystyle\sum_{i = 1}^{m_3}\lambda_{3i}v_iw_i=0.$ Hence, $\alpha_3=\displaystyle\sum_{i = 1}^{m_3}\lambda_{3i}\tau_1\tau_2(v_i^qw_i+v_iw_i^q)\gamma^4$.
Since $\alpha_2+\alpha_3=0$, we know that $\alpha_2=\alpha_3=0.$
 Then we have proved that $UV+U^2+V^2=UV\oplus U^2\oplus V^2$.

   Secondly, we prove that $U,V$ satisfy (2) in Lemma \ref{lem3.1}.
   For any nonzero elements $\beta_1=\delta_1a+\delta_2a^q\gamma^2, \beta_2=\delta_1b+\delta_2b^q\gamma^2\in U$ and $\beta_3=\tau_1u\gamma+\tau_2u^q\gamma^3, \beta_4=\tau_1v\gamma+\tau_2v^q\gamma^3\in V.$
   If $\beta_1\beta_3=\beta_2\beta_4$, that is\\
   \centerline{$(\delta_1a+\delta_2a^q\gamma^2)(\tau_1u\gamma+\tau_2u^q\gamma^3)=(\delta_1b+\delta_2b^q\gamma^2)(\tau_1v\gamma+\tau_2v^q\gamma^3)$,}
    then we have
   $\delta_1a\tau_1u=\delta_1b\tau_1v$, which means $au=bv$. Hence, $\frac{a}{b}=\frac{v}{u}=\lambda\in\mathbb{F}_{q^k}^*\cap\mathbb{F}_{q^l}^*.$ Since $\gcd(k,l)=1,$ then $\frac{\beta_1}{\beta_2}=\frac{\beta_4}{\beta_3}=\lambda\in\mathbb{F}_q^*.$

   By Lemma \ref{lem3.1}, we reach the conclusion of this theorem.
  \end{proof}
\end{thm}

\begin{remark}
  When we change the positions of $a$ and $a^q$ or $u$ and $u^q$ in Theorem \ref{thm4.7},
  new Sidon spaces can be obtained. All Sidon spaces constructed below satisfy this property.
\end{remark}

\begin{thm}
  For three positive integers $n,k,l$ and $kl\mid n,$ $\gcd(k,l)=1.$
  Let $\gamma\in\mathbb{F}_{q^n}^*$ be a root of an irreducible polynomial of degree $\frac{n}{kl}> 6$
  over $\mathbb{F}_{q^{kl}}$. Let \\
  \centerline{$\mathcal{U}=\{\delta_1 a+\delta_2 a^q\gamma+\tau_1 u\gamma^2+ \tau_2 u^q\gamma^3\mid a\in \mathbb{F}_{q^k},u\in \mathbb{F}_{q^l}\},$}
  where $\delta_i\in\mathbb{F}_{q^k}^*, \tau_i\in\mathbb{F}_{q^l}^*, i=1,2.$ Then $\mathcal{U}$
  is a Sidon space with dimension $k+l$.

  \begin{proof}
    The proof is similar to the proof of Theorem \ref{thm4.7}.
  \end{proof}
\end{thm}

\begin{remark}
 When $\delta_1=\delta_2=\tau_1=\tau_2=1$, the result of this theorem is the same as the Theorem 3.3 in \cite{NYW}. We give another proof here, which
  is different from the proof in \cite{NYW}.
\end{remark}

 We construct several Sidon spaces with the element $\gamma$ in above theorems and corollaries. It is our intention to
construct new Sidon spaces with much more distinct elements in the following theorem.
We first construct a Sidon space which is the sum of $\mathbb{F}_{q}$ and a nontrivial
Sidon space constructed by two distinct elements $\gamma$ and $\xi$.

\begin{thm}
  For three positive integers $n,k,m$ and $k\mid m\mid n.$  Let $\gamma\in\mathbb{F}_{q^m}^*$ be a root of an irreducible polynomial of degree $\frac{m}{k}> 2$
   over $\mathbb{F}_{q^{k}}$, and $\xi\in\mathbb{F}_{q^n}^*$ be a root of an irreducible polynomial of degree $\frac{n}{m}> 2$
   over $\mathbb{F}_{q^{m}}$. Let \\
   \centerline{$\mathcal{U}=\{a+\delta_1 u\gamma+\delta_2 u^{q}\xi\mid a\in\mathbb{F}_q, u\in \mathbb{F}_{q^k}\},$}
   where $\delta_i\in\mathbb{F}_{q^k}^*,i=1,2.$ Then $\mathcal{U}$ is a Sidon space with dimension $k+1$.
   \begin{proof}

   We assume that $U=\{\delta_1 u\gamma+\delta_2 u^q\xi\mid u\in\mathbb{F}_{q^k}\}.$ It is easy to verify that $U$ is a Sidon space.
   Firstly, we prove that these two Sidon spaces $\mathbb{F}_q,U$ satisfy (1) in Lemma \ref{lem3.1}.
   Let
   \begin{center}
    $\begin{aligned}
      \alpha_1&=a\in\mathbb{F}_q,\\
      \alpha_2&=\delta_1 u\gamma+\delta_2 u^q\xi\in U,\\
      \alpha_3&=\displaystyle\sum_{i = 1}^{m}\lambda_{i}[\delta_1^2 v_iw_i\gamma^2+\delta_1\delta_2(v_i^qw_i+v_iw_i^q)\gamma\xi+\delta_2^2 v_i^qw_i^q\xi^2]\in U^2,
    \end{aligned}$
   \end{center}
  where $a,\lambda_{i}\in\mathbb{F}_q, u,v_i,w_i\in \mathbb{F}_{q^k}.$

  Since $\frac{n}{m}>2$ and $\frac{m}{k}>2$, we know that \\
  \centerline{$\{1,\gamma,\xi,\gamma^2,\xi^2,\gamma\xi,\gamma^2\xi,\gamma\xi^2,\gamma^2\xi^2\}$}
   are linearly independent over $\mathbb{F}_{q^k}$.
  Hence, if $\alpha_1+\alpha_2+\alpha_3=0$, we have $\alpha_1=\alpha_2=\alpha_3=0.$
  Secondly, we prove that $U,\mathbb{F}_q$ satisfy (2) in Lemma \ref{lem3.1}.
  For any nonzero elements $\delta_1 u\gamma+\delta_2 u^q\xi, \delta_1 v\gamma+\delta_2 v^q\xi\in U$ and $a,b\in \mathbb{F}_q^*.$
  If\\
   \centerline{$a(\delta_1 u\gamma+\delta_2 u^q\xi)=b(\delta_1 v\gamma+\delta_2 v^q\xi)$,}
    then we have
  $\delta_1 au\gamma=\delta_1 bv\gamma$, which means $au=bv.$ Hence, $\frac{u}{v}=\frac{b}{a}=\frac{\delta_1 u\gamma+\delta_2 u^q\xi}{\delta_1 v\gamma+\delta_2v^q\xi}=\lambda\in\mathbb{F}_{q}^*.$

  By Lemma \ref{lem3.1}, we draw a conclusion that $\mathcal{U}$ is a Sidon space.
 \end{proof}
 \end{thm}

 \begin{remark}
  When we change the positions of $\gamma$ and $\xi$, new Sidon space can be obtained.
 \end{remark}

 In the following theorem, we construct a Sidon space which is the sum of two nontrivial Sidon spaces that are constructed by two distinct elements $\gamma$ and $\xi$.

\begin{thm}\label{thm4.13}
  For three positive integers $n,k,m$ and $k\mid m\mid n.$
  Let $\gamma\in\mathbb{F}_{q^n}^*$ be a root of an irreducible polynomial of degree $\frac{n}{k}> 2$
  over $\mathbb{F}_{q^{k}}$, and $\xi\in\mathbb{F}_{q^n}^*$ be a root of an irreducible polynomial of degree $\frac{n}{m}> 4$
  over $\mathbb{F}_{q^{m}}$. Let \\
  \centerline{$\mathcal{U}=\{\delta_1 a+\delta_2 a^q\gamma+\tau_1 u\xi+ \tau_2 u^q\xi^2\mid a,u \in \mathbb{F}_{q^k}\},$}
  where $\delta_i,\tau_i\in\mathbb{F}_{q^k}^*, i=1,2.$ Then $\mathcal{U}$
  is a Sidon space with dimension $2k$.

  \begin{proof}
    We assume that $U=\{\delta_1 a+\delta_2 a^q\gamma\mid a\in \mathbb{F}_{q^k}\}, V=\{\tau_1 u\xi+\tau_2 u^q\xi^2\mid u\in \mathbb{F}_{q^k}\}$, then $U,V$ are all Sidon spaces.
    Firstly, we prove that $U,V$ satisfy (1) in Lemma \ref{lem3.1}.
    Let
   \begin{center}
    $\begin{aligned}
   \alpha_1&=\displaystyle\sum_{i = 1}^{m_1}\lambda_{1i}[\delta_1\tau_1 a_iu_i\xi+\delta_2\tau_1a_i^qu_i\gamma\xi+\delta_1\tau_2a_iu_i^q\xi^2+\delta_2\tau_2a_i^qu_i^q\gamma\xi^2]\in UV,\\
   \alpha_2&=\displaystyle\sum_{i = 1}^{m_2}\lambda_{2i}[\delta_1^2b_ic_i+\delta_1\delta_2(b_i^qc_i+b_ic_i^q)\gamma+\delta_2^2b_i^qc_i^q\gamma^2]\in U^2, \\
   \alpha_3&=\displaystyle\sum_{i = 1}^{m_3}\lambda_{3i}[\tau_1^2v_iw_i\xi^2+\tau_1\tau_2(v_i^qw_i+v_iw_i^q)\xi^3+\tau_2^2v_i^qw_i^q\xi^4]\in V^2,
    \end{aligned}$
   \end{center}
   where $\lambda_{ij}\in\mathbb{F}_q, a_i,b_i,c_i, u_i,v_i,w_i\in \mathbb{F}_{q^k}.$
   Since $\frac{n}{m}>4$ and $\frac{m}{k}>2$, we know that $\{\gamma^i \xi^j \mid 0\leq i\leq 2, 0\leq j\leq 4 \}$ are linearly independent over $\mathbb{F}_{q^k}$.
Hence, if $\alpha_1+\alpha_2+\alpha_3=0$, then we have $\alpha_1=\alpha_2=\alpha_3=0.$

   Secondly, we prove that $U,V$ satisfy (2) in Lemma \ref{lem3.1}.
   For any nonzero elements $\beta_1=\delta_1a+\delta_2a^q\gamma, \beta_2=\delta_1b+\delta_2b^q\gamma\in U$ and $\beta_3=\tau_1u\xi+\tau_2u^q\xi^2, \beta_4=\tau_1v\xi+\tau_2v^q\xi^2\in V.$
   If $\beta_1\beta_3=\beta_2\beta_4$, that is\\
   \centerline{$(\delta_1a+\delta_2a^q\gamma)(\tau_1u\xi+\tau_2u^q\xi^2)=(\delta_1b+\delta_2b^q\gamma)(\tau_1v\xi+\tau_2v^q\xi^2)$,}
    then we have
   $\delta_1a\tau_1u\xi=\delta_1b\tau_1v\xi$ and $\delta_2a^q\gamma\tau_1u\xi=\delta_2b^q\gamma\tau_1v\xi$, hence $au=bv, a^qu=b^qv$. Therefore, $\frac{a}{b}=\frac{v}{u}=\frac{a^q}{b^q},$ which means $\frac{a}{b}=\frac{v}{u}=\lambda\in\mathbb{F}_q^*$. Thus,
$\frac{\beta_1}{\beta_2}=\frac{\beta_4}{\beta_3}=\lambda\in\mathbb{F}_q^*$.

By Lemma \ref{lem3.1}, we conclude that $\mathcal{U}$ is a Sidon space.
  \end{proof}
\end{thm}

\begin{thm}
  For three positive integers $n,k,m$ and $k\mid m\mid n.$
  Let $\gamma\in\mathbb{F}_{q^n}^*$ be a root of an irreducible polynomial of degree $\frac{n}{k}> 4$
  over $\mathbb{F}_{q^{k}}$, and $\xi\in\mathbb{F}_{q^n}^*$ be a root of an irreducible polynomial of degree $\frac{n}{m}> 2$
  over $\mathbb{F}_{q^{m}}$. Let \\
  \centerline{$\mathcal{U}=\{\delta_1 a+\delta_2 a^q\xi+\tau_1 u\gamma+ \tau_2 u^q\gamma^2\mid a,u \in \mathbb{F}_{q^k}\},$}
  where $\delta_i,\tau_i\in\mathbb{F}_{q^k}^*, i=1,2.$ Then $\mathcal{U}$
  is a Sidon space with dimension $2k$.
  \begin{proof}
    The proof is similar to the proof of Theorem \ref{thm4.13}.
  \end{proof}
\end{thm}

\subsection{Constructing Sidon spaces through three Sidon spaces}

In the following  theorem, we construct a Sidon space that is the sum of
$\mathbb{F}_{q}$ and two other nontrivial Sidon spaces.

\begin{thm}\label{thm4.15}
  For three positive integer $n,k,l$ and $kl\mid n,$ $\gcd(k,l)=1.$
  Let $\gamma\in\mathbb{F}_{q^n}^*$ be a root of an irreducible polynomial of degree $\frac{n}{kl}> 8$
  over $\mathbb{F}_{q^{kl}}$. Let \\
  \centerline{$\mathcal{U}=\{a+\delta_1u\gamma+\delta_2u^q\gamma^2+ \tau_1e\gamma^3+\tau_2e^q\gamma^4 \mid a\in \mathbb{F}_{q},u\in \mathbb{F}_{q^l},e\in\mathbb{F}_{q^k}\}$,}
  where $\delta_i\in\mathbb{F}_{q^l}^*, \tau_i\in\mathbb{F}_{q^k}^*, i=1,2.$
  Then $\mathcal{U}$ is a Sidon space with dimension $k+l+1$.

  \begin{proof}
    We assume that $U=\{\delta_1 u\gamma+\delta_2 u^q\gamma^2\mid u\in\mathbb{F}_{q^l}\}, V=\{\tau_1 e\gamma^3+\tau_2 e^q\gamma^4\mid e\in\mathbb{F}_{q^k}\}.$ Then $U,V$ are all Sidon spaces.
 Firstly, we prove that these three Sidon spaces $\mathbb{F}_q,U,V$ satisfy (1) in Theorem \ref{thm3.2}.
 Let
 \begin{center}
  $\begin{aligned}
    \alpha_1&=\displaystyle\sum_{i = 1}^{m_1}\lambda_{1i}[\delta_1\tau_1 v_if_i\gamma^4+(\delta_2\tau_1 v_i^qf_i+\delta_1\tau_2v_if_i^q)\gamma^5+\delta_2\tau_2 v_i^qf_i^q\gamma^6]\in UV,\\
    \alpha_2&=\displaystyle\sum_{i = 1}^{m_2}\lambda_{2i}[\delta_1^2 x_iy_i\gamma^2+\delta_1\delta_2(x_i^qy_i+x_iy_i^q)\gamma^3+\delta_2^2 x_i^qy_i^q\gamma^4]\in U^2,\\
    \alpha_3&=\displaystyle\sum_{i = 1}^{m_3}\lambda_{3i}[\tau_1^2 g_ih_i\gamma^6+\tau_1\tau_2 (g_i^qh_i+g_ih_i^q)\gamma^7+\tau_2^2 g_i^qh_i^q\gamma^8]\in V^2,\\
    \alpha_4&=a\in\mathbb{F}_q,\\
    \alpha_5&=\delta_1 u\gamma+\delta_2 u^q\gamma^2\in U,\\
    \alpha_6&=\tau_1 e\gamma^3+\tau_2 e^q\gamma^4\in V,
  \end{aligned}$
 \end{center}
where $\lambda_{ij}\in\mathbb{F}_q, u,v_i,x_i,y_i\in \mathbb{F}_{q^l}, e,f_i,g_i,h_i\in \mathbb{F}_{q^k}.$

If $\displaystyle\sum_{i = 1}^{6}\alpha_i=0$, it is easy to know $\alpha_4=0.$ Then we have
 $\displaystyle\sum_{i = 1}^{m_3}\lambda_{3i}\tau_2^2g_i^qh_i^q\gamma^8=0$ and $\displaystyle\sum_{i = 1}^{m_3}\lambda_{3i}\tau_1\tau_2(g_i^qh_i+g_ih_i^q)\gamma^7=0$. Thus, $\displaystyle\sum_{i = 1}^{m_3}\lambda_{3i}g_i^qh_i^q=0,$ this means
 $\displaystyle\sum_{i = 1}^{m_3}\lambda_{3i}g_ih_i=0.$ Therefore, we have $\alpha_3=0$.
 It is easy to know that $\displaystyle\sum_{i = 1}^{m_1}\lambda_{1i}\delta_2\tau_2 v_i^qf_i^q\gamma^6=0$ and $\displaystyle\sum_{i = 1}^{m_1}\lambda_{1i}(\tau_1\delta_2v_i^qf_i+\tau_2\delta_1v_if_i^q)\gamma^5=0,$ which leads to
 $\displaystyle\sum_{i = 1}^{m_1}\lambda_{1i}v_if_i=0.$ Thus, we have $\alpha_1=0.$
 Now we consider the element $\alpha_5,$ then we have $u=u^q=0$, hence $\alpha_5=0$.
 We know that $\displaystyle\sum_{i = 1}^{m_2}\lambda_{2i}\delta_1^2 x_iy_i\gamma^2=0$, which
 means $\displaystyle\sum_{i = 1}^{m_2}\lambda_{2i}x_iy_i=0$. This leads to $\displaystyle\sum_{i = 1}^{m_2}\lambda_{2i}x_i^qy_i^q=0$, then
 we have $\alpha_2=\displaystyle\sum_{i = 1}^{m_2}\lambda_{2i}\delta_1\delta_2(x_i^qy_i+x_iy_i^q)\gamma^3.$ Since $\alpha_2+\alpha_6=0,$
 it is easy to know that $e^q=e=0$ Hence, $\alpha_6=0=\alpha_2.$ Then we have proved
 that $UV+U+V+U^2+V^2+\mathbb{F}_q=UV\oplus U\oplus V\oplus U^2\oplus V^2\oplus \mathbb{F}_q$.

Secondly, we prove that $U,V,\mathbb{F}_q$ satisfy (2) in Theorem \ref{thm3.2}.
For any nonzero elements $\beta_1=\delta_1 u\gamma+\delta_2 u^q\gamma^2, \beta_2=\delta_1 v\gamma+\delta_2 v^q\gamma^2\in U$ and $\beta_3=\tau_1 e\gamma^3+\tau_2 e^q\gamma^4, \beta_4=\tau_1 f\gamma^3+\tau_2 f^q\gamma^4\in V.$
If $\beta_1\beta_3=\beta_2\beta_4$, that is\\
 \centerline{$(\delta_1 u\gamma+\delta_2 u^q\gamma^2)(\tau_1 e\gamma^3+\tau_2 e^q\gamma^4)=(\delta_1 v\gamma+\delta_2 v^q\gamma^2)(\tau_1 f\gamma^3+\tau_2 f^q\gamma^4)$,}
  then we have
$\delta_1 u\tau_1 e=\delta_1 v\tau_1 f$, which means $ue=vf.$ Hence, $\frac{u}{v}=\frac{f}{e}=\lambda\in\mathbb{F}_{q^k}^* \cap \mathbb{F}_{q^l}^*.$ Since $\gcd(k,l)=1,$ then
$\frac{\beta_1}{\beta_2}=\frac{\beta_4}{\beta_3}=\lambda\in\mathbb{F}_q^*.$
Other cases can be proved similarly.

By Theorem \ref{thm3.2}, we obtain that $\mathcal{U}$ is a Sidon space.
  \end{proof}

\end{thm}

In the following theorem, we construct a Sidon space which is the sum of three nontrivial Sidon spaces.

\begin{thm}\label{thm4.16}
  For four positive integers $n,k,l,s$ with $kls\mid n,$ and $k,l,s$ are pairwise relatively prime integers.
  Let $\gamma\in\mathbb{F}_{q^n}^*$ be a root of an irreducible polynomial of degree $\frac{n}{kls}> 12$
  over $\mathbb{F}_{q^{kls}}$. Let \\
  \centerline{$\mathcal{U}=\{\delta_1 a+\delta_2 a^q\gamma+ \tau_1 u\gamma^2+\tau_2 u^q\gamma^3+\eta_1 e\gamma^5+\eta_2 e^q\gamma^6 \mid a\in \mathbb{F}_{q^l},u\in \mathbb{F}_{q^k},e\in\mathbb{F}_{q^s}\},$}
  where $\delta_i\in\mathbb{F}_{q^l}^*, \tau_i\in\mathbb{F}_{q^k}^*, \eta_i\in\mathbb{F}_{q^s}^*, i=1,2$. Then $\mathcal{U}$ is a Sidon space with dimension $k+l+s$.

  \begin{proof}
 We assume that $U=\{\delta_1 a+\delta_2a^q\gamma\mid a\in\mathbb{F}_{q^l}\}, V=\{\tau_1 u\gamma^2+\tau_2u^q\gamma^3\mid u\in\mathbb{F}_{q^k}\}, W=\{\eta_1 e\gamma^5+\eta_2 e^q\gamma^6\mid e\in\mathbb{F}_{q^s}\}.$ Then $U,V,W$ are all Sidon spaces.
 Firstly, we prove that $U,V,W$ satisfy (1) in Theorem \ref{thm3.2}. Let
\begin{center}
 $\begin{aligned}
\alpha_1&=\displaystyle\sum_{i = 1}^{m_1}\lambda_{1i}[\delta_1\tau_1 a_iu_i\gamma^2+(\delta_2\tau_1 a_i^qu_i+\delta_1\tau_2 a_iu_i^q)\gamma^3+\delta_2\tau_2 a_i^qu_i^q\gamma^4]\in UV,\\
\alpha_2&=\displaystyle\sum_{i = 1}^{m_2}\lambda_{2i}[\tau_1\eta_1 e_iv_i\gamma^7+(\tau_1\eta_2 e_i^qv_i+\tau_2\eta_1 e_iv_i^q)\gamma^8+\tau_2\eta_2 e_i^qv_i^q\gamma^9]\in VW, \\
\alpha_3&=\displaystyle\sum_{i = 1}^{m_3}\lambda_{3i}[\delta_1\eta_1 b_if_i\gamma^5+(\delta_2\eta_1 b_i^qf_i+\delta_1\eta_2 b_if_i^q)\gamma^6+\delta_2\eta_2 b_i^qf_i^q\gamma^7]\in UW,\\
\alpha_4&=\displaystyle\sum_{i = 1}^{m_4}\lambda_{4i}[\delta_1^2 c_id_i+\delta_1\delta_2(c_i^qd_i+c_id_i^q)\gamma+\delta_2^2c_i^qd_i^q\gamma^2]\in U^2,\\
\alpha_5&=\displaystyle\sum_{i = 1}^{m_5}\lambda_{5i}[\tau_1^2 x_iy_i\gamma^4+\tau_1\tau_2(x_i^qy_i+x_iy_i^q)\gamma^5+\tau_2^2x_i^qy_i^q\gamma^6]\in V^2,
\end{aligned}$
\end{center}
\begin{center}
 $\begin{aligned}
\alpha_6&=\displaystyle\sum_{i = 1}^{m_6}\lambda_{6i}[\eta_1^2 g_ih_i\gamma^{10}+\eta_1\eta_2(g_i^qh_i+g_ih_i^q)\gamma^{11}+\eta_2^2g_i^qh_i^q\gamma^{12}]\in W^2,
 \end{aligned}$
\end{center}
where $\lambda_{ij}\in\mathbb{F}_q, a_i,b_i,c_i,d_i\in \mathbb{F}_{q^l}, u_i,v_i,x_i,y_i\in \mathbb{F}_{q^k}, e_i,f_i,g_i,h_i\in \mathbb{F}_{q^s}.$

If $\displaystyle\sum_{i = 1}^{6}\alpha_i=0$, it is easy to know $\alpha_6=0.$ Then we have
 $\displaystyle\sum_{i = 1}^{m_2}\lambda_{2i}\tau_2\eta_2 e_i^qv_i^q\gamma^9=0$ and $\displaystyle\sum_{i = 1}^{m_2}\lambda_{2i}(\tau_1\eta_2 e_i^qv_i+\tau_2\eta_1 e_iv_i^q)\gamma^8=0$. Thus, $\displaystyle\sum_{i = 1}^{m_2}\lambda_{2i}e_i^qv_i^q=\displaystyle\sum_{i = 1}^{m_2}\lambda_{2i}^q e_i^qv_i^q=0,$ this means
 $\displaystyle\sum_{i = 1}^{m_2}\lambda_{2i}e_iv_i=0.$ Therefore, we have $\alpha_2=0$.
 It is easy to know that $\displaystyle\sum_{i = 1}^{m_4}\lambda_{4i}\delta_1^2c_id_i=0$ and $\displaystyle\sum_{i = 1}^{m_4}\lambda_{4i}\delta_1\delta_2(c_i^qd_i+c_id_i^q)\gamma=0,$ which leads to
 $\displaystyle\sum_{i = 1}^{m_4}\lambda_{4i}c_i^qd_i^q=0.$ Thus, we have $\alpha_4=0.$
 Now we consider the element $\alpha_1,$
 we have $\displaystyle\sum_{i = 1}^{m_1}\lambda_{1i}\delta_1 \tau_1 a_iu_i\gamma^2$\\$=0$
 and $\displaystyle\sum_{i = 1}^{m_1}\lambda_{1i}(\delta_2\tau_1 a_i^qu_i+\delta_1\tau_2 a_iu_i^q)\gamma^3=0,$ which leads to $\displaystyle\sum_{i = 1}^{m_1}\lambda_{1i}a_i^qu_i^q=0$.
 Therefore, $\alpha_1=0$. Since $\alpha_3+\alpha_5=0,$ we know that $\displaystyle\sum_{i = 1}^{m_3}\lambda_{3i}\delta_2\eta_2 b_i^qf_i^q\gamma^7=0$, which
 means $\displaystyle\sum_{i = 1}^{m_3}\lambda_{3i}b_if_i=0$. Then
 we have $\alpha_3=\displaystyle\sum_{i = 1}^{m_3}\lambda_{3i}(\delta_2\eta_1 b_i^qf_i+\delta_1\eta_2 b_if_i^q)\gamma^6.$ When we turn to
 $\alpha_5$, it is easy to know that $\displaystyle\sum_{i = 1}^{m_5}\lambda_{5i}\tau_1^2 x_iy_i\gamma^4=0$ and $\displaystyle\sum_{i = 1}^{m_5}\lambda_{5i}\tau_1\tau_2(x_i^qy_i+x_iy_i^q)\gamma^5=0$, which leads to
 $\displaystyle\sum_{i = 1}^{m_5}\lambda_{5i}x_i^qy_i^q=0$. Hence, $\alpha_5=0=\alpha_3.$ Then we have proved
 that $UV+VW+UW+U^2+V^2+W^2=UV\oplus VW\oplus UW\oplus U^2\oplus V^2\oplus W^2$.

Secondly, we prove that $U,V,W$ satisfy (2) in Theorem \ref{thm3.2}.
For any nonzero elements $\beta_1=\delta_1 a+\delta_2 a^q\gamma, \beta_2=\delta_1 b+\delta_2 b^q\gamma\in U$ and $\beta_3=\tau_1 u\gamma^2+\tau_2 u^q\gamma^3, \beta_4=\tau_1 v\gamma^2+\tau_2 v^q\gamma^3\in V.$
If $\beta_1\beta_3=\beta_2\beta_4$, that is\\
\centerline{$(\delta_1 a+\delta_2a^q\gamma)(\tau_1 u\gamma^2+\tau_2 u^q\gamma^3)=(\delta_1 b+\delta_2 b^q\gamma)(\tau_1 v\gamma^2+\tau_2 v^q\gamma^3)$,}
 then we have
$\delta_1 a\tau_1 u=\delta_1 b\tau_1 v$, which means $au=bv$. Hence, $\frac{a}{b}=\frac{v}{u}=\lambda\in\mathbb{F}_{q^k}^*\cap\mathbb{F}_{q^l}^*.$ Since $\gcd(k,l)=1,$ then $\frac{\beta_1}{\beta_2}=\frac{\beta_4}{\beta_3}=\lambda\in\mathbb{F}_q^*$.
Other cases can be proved similarly.

By Theorem \ref{thm3.2}, we conclude that $\mathcal{U}$ is a Sidon space.
  \end{proof}
\end{thm}

\begin{thm}
  For four positive integers $n,k,l,s$ with $kls\mid n,$ and $k,l,s$ are pairwise relatively prime integers.
  Let $\gamma\in\mathbb{F}_{q^n}^*$ be a root of an irreducible polynomial of degree $\frac{n}{kls}> 12$
  over $\mathbb{F}_{q^{kls}}$. Let \\
  \centerline{$\mathcal{U}=\{\delta_1 a+\delta_2 a^q\gamma+\tau_1 u\gamma^3+\tau_2 u^q\gamma^4+\eta_1 e\gamma^5+\eta_2 e^q\gamma^6 \mid a\in \mathbb{F}_{q^l},u\in \mathbb{F}_{q^k},e\in\mathbb{F}_{q^s}\},$}
  where $\delta_i\in\mathbb{F}_{q^l}^*, \tau_i\in\mathbb{F}_{q^k}^*, \eta_i\in\mathbb{F}_{q^s}^*, i=1,2$.
 Then $\mathcal{U}$ is a Sidon space with dimension $k+l+s$.

\begin{proof}

 The proof is similar to the proof of Theorem \ref{thm4.16}.
\end{proof}
\end{thm}

In above theorems, we have constructed several Sidon spaces with the element $\gamma$.
In the following theorem, we will construct a Sidon space which is the sum of three nontrivial Sidon spaces that are constructed by
three distinct elements  respectively.

\begin{thm}\label{thm4.18}
  For four positive integers $n,k,m,r$ and $k\mid m \mid r\mid n.$
  Let $\gamma\in\mathbb{F}_{q^m}^*$ be a root of an irreducible polynomial of degree $\frac{m}{k}> 2$
  over $\mathbb{F}_{q^{k}}$, and $\xi\in\mathbb{F}_{q^r}^*$ be a root of an irreducible polynomial of degree $\frac{r}{m}> 4$
  over $\mathbb{F}_{q^{m}}$, and $\zeta\in\mathbb{F}_{q^n}^*$ be a root of an irreducible polynomial of degree $\frac{n}{r}> 4$
  over $\mathbb{F}_{q^{r}}$. Let \\
  \centerline{$\mathcal{U}=\{\delta_1 a+\delta_2 a^q\gamma+ \tau_1 u\xi+\tau_2 u^q\xi^2+\eta_1 e\zeta+\eta_2 e^q\zeta^2 \mid a,u,e\in \mathbb{F}_{q^k}\},$}
  where $\delta_i, \tau_i, \eta_i \in\mathbb{F}_{q^k}^*, i=1,2.$ Then $\mathcal{U}$ is a Sidon space with dimension $3k$.

  \begin{proof}
 We assume that $U=\{\delta_1 a+\delta_2a^q\gamma\mid a\in\mathbb{F}_{q^k}\}, V=\{\tau_1 u\xi+\tau_2u^q\xi^2\mid u\in\mathbb{F}_{q^k}\}, W=\{\eta_1 e\zeta+\eta_2 e^q\zeta^2\mid e\in\mathbb{F}_{q^k}\}.$ Then $U,V,W$ are all Sidon spaces.
 Firstly, we prove that $U,V,W$ satisfy (1) in Theorem \ref{thm3.2}.
 Let
\begin{center}
 $\begin{aligned}
\alpha_1&=\displaystyle\sum_{i = 1}^{m_1}\lambda_{1i}[\delta_1\tau_1 a_iu_i\xi+\delta_2\tau_1 a_i^qu_i\gamma\xi+\delta_1\tau_2 a_iu_i^q\xi^2+\delta_2\tau_2 a_i^qu_i^q\gamma\xi^2]\in UV,\\
\alpha_2&=\displaystyle\sum_{i = 1}^{m_2}\lambda_{2i}[\tau_1\eta_1 e_iv_i\xi\zeta+\tau_1\eta_2 e_i^qv_i\xi\zeta^2+\tau_2\eta_1 e_iv_i^q\zeta\xi^2+\tau_2\eta_2 e_i^qv_i^q\zeta^2\xi^2]\in VW, \\
\alpha_3&=\displaystyle\sum_{i = 1}^{m_3}\lambda_{3i}[\delta_1\eta_1 b_if_i\zeta+\delta_2\eta_1 b_i^qf_i\gamma\zeta+\delta_1\eta_2 b_if_i^q\zeta^2+\delta_2\eta_2 b_i^qf_i^q\gamma\zeta^2]\in UW,\\
\alpha_4&=\displaystyle\sum_{i = 1}^{m_4}\lambda_{4i}[\delta_1^2 c_id_i+\delta_1\delta_2(c_i^qd_i+c_id_i^q)\gamma+\delta_2^2c_i^qd_i^q\gamma^2]\in U^2,\\
\alpha_5&=\displaystyle\sum_{i = 1}^{m_5}\lambda_{5i}[\tau_1^2 x_iy_i\xi^2+\tau_1\tau_2(x_i^qy_i+x_iy_i^q)\xi^3+\tau_2^2x_i^qy_i^q\xi^4]\in V^2,\\
\alpha_6&=\displaystyle\sum_{i = 1}^{m_6}\lambda_{6i}[\eta_1^2 g_ih_i\zeta^{2}+\eta_1\eta_2(g_i^qh_i+g_ih_i^q)\zeta^{3}+\eta_2^2g_i^qh_i^q\zeta^{4}]\in W^2,
 \end{aligned}$
\end{center}
where $\lambda_{ij}\in\mathbb{F}_q, a_i,b_i,c_i,d_i, u_i,v_i,x_i,y_i, e_i,f_i,g_i,h_i\in \mathbb{F}_{q^k}.$
Since $\frac{n}{r}>4,\frac{r}{m}>4$ and $\frac{m}{k}>2$, we know that $\{\gamma^i \xi^j\zeta^l\mid 0\leq i\leq 2, 0\leq j,l\leq 4 \}$ are linearly independent over $\mathbb{F}_{q^k}$.
Hence, if $\alpha_1+\alpha_2+\alpha_3+\alpha_4+\alpha_5+\alpha_6=0$, then we have $\alpha_1=\alpha_2=\alpha_3=\alpha_4+\alpha_5=\alpha_6=0.$

Secondly, we prove that $U,V,W$ satisfy (2) in Theorem \ref{thm3.2}.
For any nonzero elements $\beta_1=\delta_1 a+\delta_2 a^q\gamma, \beta_2=\delta_1 b+\delta_2 b^q\gamma\in U$ and $\beta_3=\tau_1 u\xi+\tau_2 u^q\xi^2, \beta_4=\tau_1 v\xi+\tau_2 v^q\xi^2\in V.$
If $\beta_1\beta_3=\beta_2\beta_4$, that is\\
\centerline{$(\delta_1 a+\delta_2a^q\gamma)(\tau_1 u\xi+\tau_2 u^q\xi^2)=(\delta_1 b+\delta_2 b^q\gamma)(\tau_1 v\xi+\tau_2 v^q\xi^2)$,}
 then we have
$\delta_1 a\tau_1 u\xi=\delta_1 b\tau_1 v\xi$ and $\delta_2a^q\gamma\tau_1 u\xi=\delta_2 b^q\gamma\tau_1 v\xi$, hence $au=bv$ and $a^qu=b^qv$.
Therefore, $\frac{a}{b}=\frac{v}{u}=\frac{a^q}{b^q}=\lambda\in\mathbb{F}_q^*.$ Thus,
$\frac{\beta_1}{\beta_2}=\frac{\beta_4}{\beta_3}=\lambda\in\mathbb{F}_q^*.$
Other cases can be proved similarly.

By Theorem \ref{thm3.2}, we reach the conclusion of this theorem.
\end{proof}
\end{thm}

\begin{thm}
  For four positive integers $n,k,m,r$ and $k\mid m \mid r\mid n.$
  Let $\gamma\in\mathbb{F}_{q^m}^*$ be a root of an irreducible polynomial of degree $\frac{m}{k}> 4$
  over $\mathbb{F}_{q^{k}}$, and $\xi\in\mathbb{F}_{q^r}^*$ be a root of an irreducible polynomial of degree $\frac{r}{m}> 2$
  over $\mathbb{F}_{q^{m}}$, and $\zeta\in\mathbb{F}_{q^n}^*$ be a root of an irreducible polynomial of degree $\frac{n}{r}> 4$
  over $\mathbb{F}_{q^{r}}$. Let \\
  \centerline{$\mathcal{U}=\{\delta_1 a+\delta_2 a^q\xi+ \tau_1 u\gamma+\tau_2 u^q\gamma^2+\eta_1 e\zeta+\eta_2 e^q\zeta^2 \mid a,u,e\in \mathbb{F}_{q^k}\},$}
  where $\delta_i, \tau_i, \eta_i \in\mathbb{F}_{q^k}^*, i=1,2.$ Then $\mathcal{U}$ is a Sidon space with dimension $3k$.

  \begin{proof}
    The proof is similar to that of Theorem \ref{thm4.18}.
  \end{proof}
\end{thm}

\begin{thm}
  For four positive integers $n,k,m,r$ and $k\mid m \mid r\mid n.$
  Let $\gamma\in\mathbb{F}_{q^m}^*$ be a root of an irreducible polynomial of degree $\frac{m}{k}> 4$
  over $\mathbb{F}_{q^{k}}$, and $\xi\in\mathbb{F}_{q^r}^*$ be a root of an irreducible polynomial of degree $\frac{r}{m}> 4$
  over $\mathbb{F}_{q^{m}}$, and $\zeta\in\mathbb{F}_{q^n}^*$ be a root of an irreducible polynomial of degree $\frac{n}{r}> 2$
  over $\mathbb{F}_{q^{r}}$. Let \\
  \centerline{$\mathcal{U}=\{\delta_1 a+\delta_2 a^q\zeta+ \tau_1 u\gamma+\tau_2 u^q\gamma^2+\eta_1 e\gamma+\eta_2 e^q\gamma^2 \mid a,u,e\in \mathbb{F}_{q^k}\},$}
  where $\delta_i, \tau_i, \eta_i \in\mathbb{F}_{q^k}^*, i=1,2.$ Then $\mathcal{U}$ is a Sidon space with dimension $3k$.

  \begin{proof}
    The proof is similar to that of Theorem \ref{thm4.18}.
  \end{proof}
\end{thm}

\subsection{Combing subspace codes}

In this section, we construct new subspace code with larger size through combing two subspace codes.
Recall that $W_{q-1}$ is the set of $(q-1)$th powers of elements in $\mathbb{F}_{q^k}$,
that is $W_{q-1}=\{x^{q-1}\mid x\in \mathbb{F}_{q^k}\}$, and $\overline{W}_{q-1}=\mathbb{F}_{q^k}\backslash{W}_{q-1}$.
\begin{thm}
  For two positive integers $n,k$ and $k\mid n.$ Let $\gamma\in\mathbb{F}_{q^n}^*$ be a root of an irreducible polynomial of degree $\frac{n}{k}> 2$
  over $\mathbb{F}_{q^{k}}$. Let \\
  \centerline{$U=\{ u+u^q\gamma \mid u\in \mathbb{F}_{q^k}\}$ and $V=\{\delta v+v^q\gamma \mid v\in \mathbb{F}_{q^k}\}$,}
  where $\delta\in\overline{W}_{q-1}$. Then\\
  \centerline{$\mathcal{C}=\{\alpha U\mid \alpha\in\mathbb{F}_{q^n}^*\}\cup \{\beta V\mid \beta\in\mathbb{F}_{q^n}^*\}$}
  is a subspace code with size $\frac{2(q^n-1)}{q-1}$ and the minimum distance $2k-2.$

  \begin{proof}
   Since $U$ and $V$ are Sidon spaces, it is enough to prove that $\dim(\alpha U,V)\leq 1$, for any $\alpha\in\mathbb{F}_{q^n}^*.$
   For any nonzero elements $a=u+u^q\gamma,a'=u'+u'^q\gamma\in U,$ and
   $b=\delta v+v^q\gamma, b'=\delta v'+v'^q\gamma\in V.$ Let\\
   \centerline{$f(x)=(u+u^qx)(\delta v+v^qx)=\delta uv+(\delta vu^q+uv^q)x+u^qv^qx^2\in\mathbb{F}_{q^k}[x]$}
   and \\
  \centerline{$f'(x)=(u'+u'^qx)(\delta v'+v'^qx)=\delta u'v'+(\delta v'u'^q+u'v'^q)x+u'^qv'^qx^2\in\mathbb{F}_{q^k}[x].$}
We know that \\
\centerline{$ab=(u+u^q\gamma)(\delta v+v^q\gamma)=\delta uv+(\delta vu^q+uv^q)\gamma+u^qv^q\gamma^2$.}
Since $\frac{n}{k}> 2$, we have $\{1,\gamma,\gamma^2\}$ are linearly independent over $\mathbb{F}_{q^{k}}$.
 Hence, if $ab=a'b'$, then we have $\delta uv=\delta u'v', \delta vu^q+uv^q=\delta v'u'^q+u'v'^q, u^qv^q=u'^qv'^q$, which means $f(x)=f'(x)$.
  Then $f(x)$ and $f'(x)$ have the same roots. If $\frac{1}{u^{q-1}}=\frac{1}{u'^{q-1}}$,
  this leads to $(\frac{u}{u'})^{q-1}=1,$ which means $\frac{u}{u'}=\lambda\in\mathbb{F}_{q}^*.$
  Therefore, $\frac{a}{a'}=\frac{b'}{b}=\lambda\in\mathbb{F}_{q}^*.$
  If $\frac{1}{u^{q-1}}=\frac{1}{\delta v'^{q-1}}$, we have $\delta=(\frac{u}{v'})^{q-1}$, this is a contradiction.
  If $\frac{1}{\delta v^{q-1}}=\frac{1}{\delta v'^{q-1}}$, it is similar to the case when
  $\frac{1}{u^{q-1}}=\frac{1}{u'^{q-1}}$.
  \end{proof}
\end{thm}

\begin{exam}
  For two positive integers $n,k$ and $k\mid n.$ Let $\gamma\in\mathbb{F}_{3^n}^*$ be a root of an irreducible polynomial of degree $\frac{n}{k}> 2$
  over $\mathbb{F}_{3^{k}}$. Let \\
  \centerline{$U=\{ u+u^3\gamma \mid u\in \mathbb{F}_{3^k}\}$ and $V=\{v-v^3\gamma \mid v\in \mathbb{F}_{3^k}\}$,}
   Then\\
  \centerline{$\mathcal{C}=\{\alpha U\mid \alpha\in\mathbb{F}_{3^n}^*\}\cup \{\beta V\mid \beta\in\mathbb{F}_{3^n}^*\}$}
  is a subspace code with size $3^n-1$ and the minimum distance $2k-2.$

  \begin{proof}
Since $-1\neq x^2, x\in \mathbb{F}_3,$ and the rest of proof is similar to the proof of the above theorem.
  \end{proof}
\end{exam}

\section{Concluding remarks}

We present several new subspace codes via new Sidon spaces in this paper. The crucial point is that we present a sufficient condition for the sum of Sidon spaces is also a Sidon space.
We generalize some results in \cite{NYW} and \cite{RRT} and give new proof for them.
Based on the sufficient condition  given in section 3.1,
 we provide new methods to construct Sidon spaces. This is meaningful for finding new way of constructing subspace codes.
\\

\noindent\textbf{Acknowledgement.} This work was supported by NSFC (Grant No. 11871025).

\end{document}